\documentclass[10pt,letterpaper,twocolumn]{article} 

\usepackage{ol2}
\usepackage[draft]{hyperref}
\usepackage{amsmath}

\begin{document}

\twocolumn[ 

\title{Low-noise on-chip frequency conversion by four-wave-mixing Bragg scattering in SiN$_{x}$ waveguides}
\vskip -0.2in

\author{Imad Agha,$^{1,2,\dagger,\ast}$ Marcelo Davan\c{c}o,$^{1,2,\dagger}$  Bryce Thurston,$^{1}$ and Kartik Srinivasan$^{1,\ast}$}

\address{$^1$Center for Nanoscale Science and Technology, National
Institute of Standards and Technology, Gaithersburg, MD 20899\\
$^2$Maryland NanoCenter, University of Maryland, College Park, MD
20742\\
$^{\dagger}$These authors contributed equally.
$^{\ast}$e-mail:imad.agha@nist.gov; kartik.srinivasan@nist.gov\\}
\vskip -0.1in

\begin{abstract}
Low-noise, tunable wavelength-conversion through non-degenerate
four-wave mixing Bragg scattering in SiN$_x$ waveguides is
experimentally demonstrated. Finite element method simulations of
waveguide dispersion are used with the split-step Fourier method to
predict device performance. Two 1550 nm wavelength band pulsed pumps
are used to achieve tunable conversion of a 980~nm signal over a
range of 5~nm with a peak conversion efficiency of $\approx5~\%$.
The demonstrated Bragg scattering process is suitable for frequency
conversion of quantum states of light.
\end{abstract}
\ocis{350.4238, 270.0270, 130.7405} \vskip -0.3in
] 

\noindent


Four-wave mixing in optical fibers has led to parametric amplifiers,
oscillators, and wavelength convertors~\cite{ref:Agrawal_NFO}. Such
behavior has recently been shown in chip-based
devices~\cite{ref:Lin_Painter_Agrawal}, where the strong modal
confinement and large $\chi^{(3)}$ in silicon and silicon nitride
(SiN$_{x}$) waveguides (WGs) enhance the effective nonlinearity
compared to fiber, and where group velocity dispersion (GVD) can be
tailored to achieve phase matching~\cite{ref:Turner_Gaeta}. Much
focus is on the configuration in which a degenerate pump beam is
placed near the zero GVD point WG, amplifying a weak input and
simultaneously generating a symmetrically-situated idler in
frequency space. This has been used to show parametric gain in
WGs~\cite{ref:Foster_Gaeta} and frequency comb generation in
microresonators~\cite{ref:Levy_Lipson_comb,ref:Ferdous_Weiner}.
However, from the perspective of quantum frequency
conversion~\cite{ref:Kumar_OL}, there is a fundamental problem, as
signal amplification comes with amplified vacuum fluctuations,
preventing noiseless operation~\cite{ref:McKinstrie}. This process,
essential for connecting quantum systems operating at disparate
wavelengths, has been applied to single photon states through
through sum-frequency generation in a few cm-long
quasi-phase-matched WG~\cite{ref:Rakher_NPhot_2010} and
four-wave-mixing Bragg scattering (FWM-BS) in a several meter long
photonic crystal fiber~\cite{ref:McGuinnes_PRL10}. Here, we make
progress towards quantum frequency conversion in an integrated
platform by demonstrating FWM-BS in SiN$_x$ WGs. We show tunable
conversion of 980 nm signals via non-degenerate 1550 nm band pumps,
with a conversion efficiency reaching 5$~\%$.

In FWM-BS, two non-degenerate pumps at frequencies $\omega_1$
and $\omega_2$ ($\omega_1>\omega_2$) scatter photons from a signal
at $\omega_s$ to an idler at $\omega_i$~\cite{ref:Uesaka_Kazovksy}.
From conservation of energy, idlers can be produced at $\omega_i^\pm$=$\omega_s\pm(\omega_1-\omega_2)$ (Fig.
1(a)), and phase matching, effective nonlinearity, interaction
length, and pump powers determine the conversion efficiency. FWM-BS
directly transfers power from signal to idler, rather than from the
pumps to the signal and idler. This avoids excess noise associated
with parametric gain processes such as modulation interaction, which
amplify vacuum fluctuations~\cite{ref:McKinstrie}. Conversion over
both small and large wavelength separations is possible if
phase-matching can be obtained.

\begin{figure}[t]
\centerline{\includegraphics[width=8.5 cm]{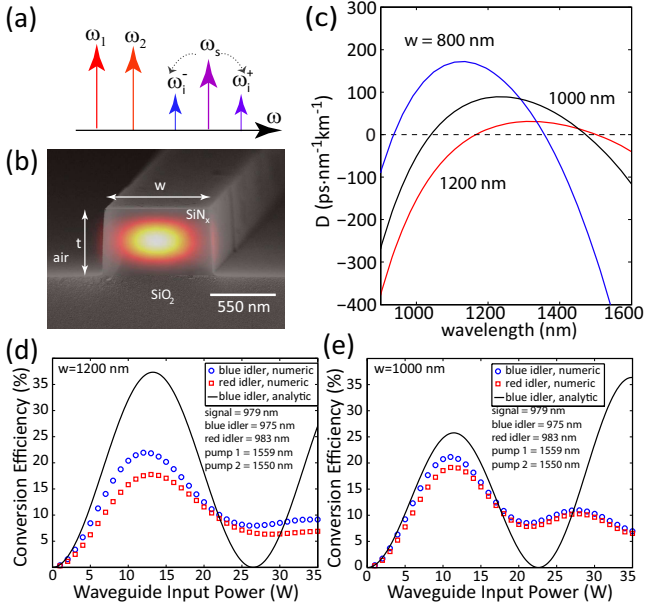}} \vskip
-0.1in \caption{(a) Schematic of FWM-BS. Pumps at $\omega_{1,2}$ and
a signal at $\omega_{s}$ create blue-shifted ($\omega_i^+$) and
red-shifted ($\omega_i^-$) idlers. (b) Fabricated SiN$_x$ WG with
superimposed fundamental TE mode profile at 1550~nm. (c) Calculated
dispersion parameter $D$ for varying $w$. (d)-(e) Numeric and
semi-analytic results for pump power dependent conversion efficiency
in 1200 nm and 1000 nm WGs. Power levels are for one pump; the two
pumps have equal power.}\label{fig:Fig1} \vskip -0.2in
\end{figure}

We are interested in quantum frequency conversion of $\approx$980 nm
signals, for eventual use with fiber-coupled quantum dot (QD) single
photon sources~\cite{ref:Davanco_WG}. Conversion is achieved using
two 1550 nm band pumps (Fig.~\ref{fig:Fig1}(a)), where the use of
far red-detuned pumps with respect to the signal avoids amplified
spontaneous emission from the pumps and potential Raman scattering,
an important noise source in fibers~\cite{ref:Gnauck}. Spontaneous
FWM is also avoided, as the process is not phase matched in our
waveguides. Together, this ensures that background-free conversion
can be achieved. As our pump wavelength separation is limited to
30~nm, the signal will be translated by at most $\approx$12 nm. Such
narrowband conversion can restore spectral indistinguishability of
independent QD single photon sources. Broader conversion ranges are
also of interest~\cite{ref:Rakher_NPhot_2010}, and a $>$100~nm
conversion range is theoretically possible in the geometries shown
here.

We calculate transverse electric polarized modes for a 550 nm tall
rectangular SiN$_x$ WG on a SiO$_2$ bottom cladding
(Fig.~\ref{fig:Fig1}(b)) over a range of width $w$ and wavelength,
allowing us to estimate dispersion relations and FWM nonlinearity.
Figure~\ref{fig:Fig1}(c) shows a plot of the dispersion parameter
$D=\frac{-2\pi c}{\lambda^2}\frac{d^2\beta}{d\omega^2}$ ($c$ is the
speed of light and $\beta$ is the WG propagation constant) for
$w$=800~nm, 1000~nm, and 1200~nm WGs, indicating that as $w$
increases, the dispersion becomes flatter and its zero point
red-shifts. Since we are targeting conversion in the 980~nm band via
1550 nm band pumps, we choose dimensions for which the dispersion
zero is close to 1200 nm~\cite{ref:Uesaka_Kazovksy}. Conversion
efficiency as a function of pump power (the pumps have equal power)
is then calculated via analytical expressions based on coupled mode
theory in the non-depleted pump
regime~\cite{ref:Uesaka_Kazovksy,ref:Agrawal_NFO} and a full-field
split-step Fourier numerical simulation \cite{ref:Agrawal_NFO},
using a WG loss of 1 dB/cm, ellipsometric measurements of the
SiN$_x$ and SiO$_2$ linear refractive indices, and a nonlinear
refractive index
$n_{2}=2.5{\times}10^{-19}$m$^{2}$W$^{-1}$~\cite{ref:Ikeda_Fainman}
that yields an effective nonlinearity parameter
$\gamma_{\text{eff}}\approx6.3$~W$^{-1}$m$^{-1}$ for$w$=1200 nm.
While the analytical solution is valid at low pump powers, it fails
to account for pump depletion and effects such as multi-frequency
Bragg scattering (due to secondary generated pumps). The split-step
Fourier simulation alleviates this since the single field launched
into the simulation includes all frequencies between the pumps and
signal/idlers, with a spectral resolution finer than the pulse
bandwidth. Moreover, it takes into account higher-order dispersion
(8 orders are included) and pulse-broadening and temporal walk-off
effects, which can be important for short pulses and wide frequency
separations. Figure~\ref{fig:Fig1}(d)-(e) shows the results for 12
mm long WGs with $w$=1200 nm and $w$=1000 nm, respectively. Both
red- and blue-detuned idlers are generated (both are nearly
phase-matched) with conversion efficiencies as high as 20$~\%$
predicted by the split-step calculation. Pump depletion and mixing
lead to the discrepancy between the split-step and analytic results
at high powers, as $>$~40$~\%$ of the pump power is consumed by pump
mixing for an input power of 10~W.

\begin{figure*}[h]
\centerline{\includegraphics[width=14cm]{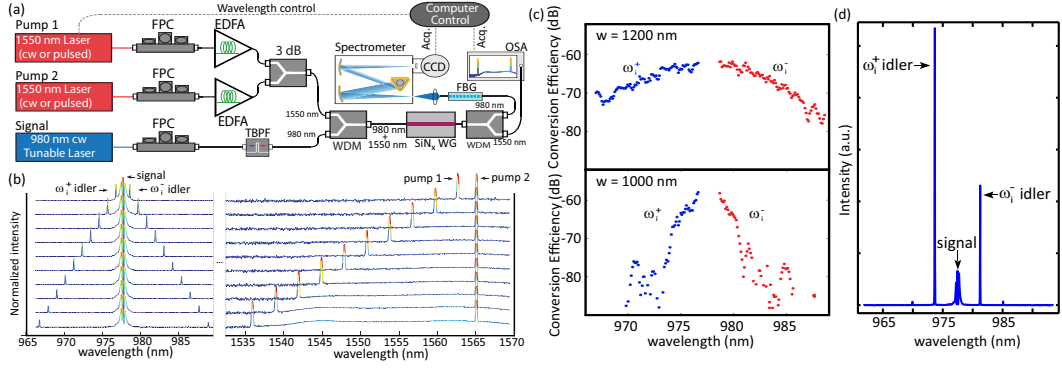}} \vskip -0.1in
\caption{(a) Experimental setup. FPC=fiber polarization controller.
(b) Simultaneous spectrometer (980 nm band) and OSA (1550 nm band)
acquired spectra of residual signal, generated idlers, and pumps 1
and 2, for varying pump 1 wavelengths. (c) Conversion efficiencies
for $\omega_i^{\pm}$ idlers as functions of idler wavelength for
$w=$1200~nm and $w=$1000~nm WGs. (d) $w$=1000 nm WG output spectrum
under pulsed (1 ns width) pumping with 6.5~W peak pump power.  The
residual signal is suppressed relative to (b) due to use of a second
FBG rejection filter.}\label{fig:Fig2} \vskip -0.05in
\end{figure*}

We fabricate 12 mm long WGs (Fig.~\ref{fig:Fig1}(b)) through a
process similar to that in Ref.~\cite{ref:Ferdous_Weiner}. Devices
are measured using the setup in Fig.~\ref{fig:Fig2}(a). To measure
conversion bandwidth, two amplified 1550 nm band continuous wave
(cw) pumps are combined with a weak signal at 977.4 nm and sent into
1000 nm and 1200 nm wide WGs via a lensed optical fiber. Light is
collected at the WG output with a lensed fiber and routed to a
wavelength division multiplexer (WDM) that separates the pumps from
the signal and idlers. The pumps are monitored on an optical
spectrum analyzer (OSA), while a grating spectrometer with a silicon
CCD measures the generated idlers and residual signal, which is
suppressed by 53 dB using a fiber Bragg grating (FBG) placed before
the spectrometer input. Pump 1 is swept between 1535 nm and 1565 nm,
while pump 2 is fixed at 1565 nm. Both FWM-BS generated $w_i^{\pm}$
idlers are visible around the signal, and move symmetrically away
from it as the separation between the pumps increases
(Fig.~\ref{fig:Fig2}(b)), in agreement with energy conservation. The
bandwidth can be deduced from Fig.~\ref{fig:Fig2}(c), where the
internal conversion efficiency $P_i/P_s$ for both idlers is plotted
($P_i$, $P_s$ are the idler and signal powers at the WG output). The
1200 nm WG has a broader conversion bandwidth than the 1000 nm WG.

To reach the high peak powers needed for more efficient
conversion, pulses from a 80 MHz repetition-rate mode-locked laser
are filtered by 1 nm wide bandpass filters at 1550 nm and 1559 nm.
The pulses, with a full-width at half-maximum
of $4.2$~ps$\pm$~1 ps, are each amplified by a 1 W erbium-doped
fiber amplifier (EDFA) and temporally overlapped by a tunable
optical-delay line before being combined with a weak 979 nm cw
signal and sent into the WG. To determine peak power while avoiding
spurious effects in the OSA, the average power is first measured at
low amplification and scaled by the duty cycle. An auxiliary SiN$_x$
WG showing efficient 3rd harmonic generation is used to calibrate
peak powers at higher amplification, due to its cubic scaling with
peak power. This allows an estimate of the maximum peak power in
the WG of $\approx~6.8$~W (accounting for $\approx$6 dB coupling
loss per facet). Keeping the pump and signal wavelengths at 1559 nm,
1550 nm, and 979 nm, the coupled power is varied between 0.5 W and 6.8 W, and the converted
idlers are measured at the WG output. The conversion efficiency,
which takes into account that the signal is cw while the pumps
are pulsed, is determined by integrating over the idler spectrum,
scaling by the pulse duty cycle, and dividing by the integrated
signal power.

\begin{figure}[t]
\centerline{\includegraphics[width=8.5cm]{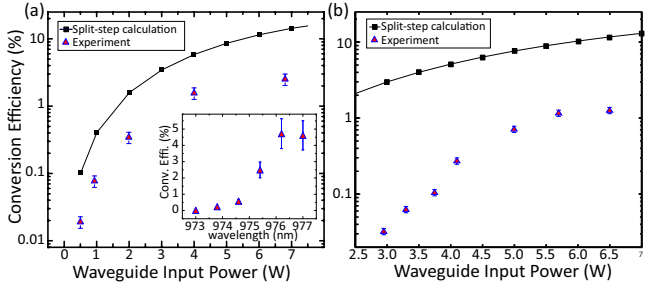}} \vskip
-0.1in \caption{Conversion efficiency vs. peak power for a 975 nm
idler (979 nm signal) in (a) a $w$=1200 nm WG using ps pump pulses,
with the inset showing the efficiency for varying idler wavelength;
(b) a $w$=1000 nm WG using ns pump pulses. Error bars are one
standard deviation values due to uncertainty in pump pulse width.
Power levels are for one pump; the two pumps have equal power. }
\vskip -0.2in \label{fig:Fig3}
\end{figure}

Figure~\ref{fig:Fig3}(a) shows conversion efficiency for the
blue-shifted idler as a function of peak pump power. The data
follows the calculated trend and reaches $\approx$~2.5~$\%$. When
the longer wavelength pump is moved to 1557 nm, where the laser
power is higher and phase-mismatch is reduced, the conversion
efficiency increases to $\approx$~5~$\%$ (inset to
Fig.~\ref{fig:Fig3}(a)). Higher efficiency may be possible with
increased pump power (Fig.~\ref{fig:Fig1}(d)), longer WGs, and more
precise dispersion tailoring.  In particular, the measured
conversion efficiency is consistently lower than predicted. While
imprecise knowledge of the effective nonlinearity and WG input power
plays a role in the discrepancy, simulations indicate that
non-optimal dispersion can be a dominant factor. An inaccurate
estimate of the WG dimensions (by $\approx$25~nm) can cause
significant changes in the predicted conversion efficiency, with a
stretching and shifting of the peaks in Fig.~\ref{fig:Fig1}(a) to
higher powers.

We next studied the 1000 nm width WG using 1 ns long pulses created
by electro-optically modulating and amplifying 1559 nm and 1550 nm
cw lasers to achieve similar peak powers as the ps pulse experiment.
This ns regime is of particular importance for wavelength conversion
of single photons from QDs, whose radiative lifetime is $\approx$1
ns. Figure~\ref{fig:Fig3}(b) shows the conversion efficiency as a
function of pump power, reaching a maximum of 1.3~$\%$, which is
around half the efficiency of the 1200 nm WG at the same peak power.
The loss of conversion and deviation from the predicted trend at
lower powers is most likely due to damage to the WGs during the
course of data accumulation (data was recorded from high to low
power), resulting from the high pulse energies.

Despite the high pump energies, no excess noise was seen in the
conversion bands (Fig.~\ref{fig:Fig2}(d)).  To confirm this, photon
counting measurements were performed. A 0.2~nm bandwidth grating
filter was spectrally aligned to the blue-detuned idler, and the
output light was detected by a Si single photon counter with the
1550~nm pump fields kept on and the 979~nm signal turned off. No
excess noise above the detector dark counts ($\approx100$~s$^{-1}$)
was measured. Thus, an input single photon source producing $10^6$
photons/s~\cite{ref:Davanco_WG} should yield a frequency-converted
flux $>10^3$ photons/s (at $>2~\%$ conversion efficiency and 12~dB
fiber-to-fiber loss), which is an order of magnitude above the
detector dark count level.

In conclusion, we have demonstrated chip-scale wavelength conversion
in a silicon nitride waveguide through the process of four-wave
mixing Bragg scattering. The background-free nature of this approach
should enable frequency conversion of quantum states of light.

We thank Houxun Miao and Rich Kasica for help with fabrication, Nanh
Van Nguyen for ellipsometer measurements, Michael Raymer for helpful
discussions, and the DARPA MESO program for partial support. I.A and
M.D. acknowledge support under the Cooperative Research Agreement
between the University of Maryland and NIST-CNST, Award
70NANB10H193. \vspace{-1.25em}


\end{document}